\documentclass{elsart}
\usepackage{epsfig,hyperref}
\usepackage{amssymb}
\usepackage{amsfonts}
\usepackage{latexsym,bm}

\long\def\comment#1{ }

\newcommand{\ts}{{\theta}}
\newcommand{\tb}{{\overline \theta}}
\newcommand{\cb}{{\overline c}}
\newcommand{\nls}{{NL$\sigma$}}
\begin{document}

\begin{frontmatter}

\parbox[]{16.6cm}{ \begin{center}

\title{A linear realization of the BRST symmetry}

\author[paris]{Matthieu Tissier\thanksref{emailmt}}
\author[montevideo]{Nicol\'as Wschebor\thanksref{emailnw}}

\address[paris]{Laboratoire de Physique Th\'eorique de la Mati\`ere Condens\'ee,
Universit\'e Pierre et Marie Curie, 4 Place Jussieu 75252 Paris CEDEX 05, France}

\address[montevideo]{ Instituto de F\'{\i}sica, Facultad de Ingenier\'{\i}a, Universidad de la Rep\'ublica, 
J.H.y Reissig 565, 11000
  Montevideo, Uruguay}

\thanks[emailmt]{tissier@lptmc.jussieu.fr}
\thanks[emailnw]{nicws@fing.edu.uy}

\date{\today}
\vspace{0.8cm}
\begin{abstract}

  In this Letter, we propose a model which is equivalent to the
  Yang-Mills theory at long distances but for which all symmetries are
  realized linearly.  On top of the gauge and Fadeev-Popov ghosts
  fields, the model presents several massive fields, all of which can
  be merged in a unique vector field with support in a curved
  superspace. The equivalence of this model with the gauge-fixed
  Yang-Mills theory is shown to result from the decoupling of these
  extra massive modes at small momenta. Most of the symmetries of
  Yang-Mills theory, including the famous BRST symmetry, can be
  interpreted in this way as isometries of the superspace.
\end{abstract}
\end{center}}

\end{frontmatter}

\section{Introduction}

Nonlinear realizations of symmetries are notoriously difficult to
handle. In a linear realization, the effective action (the generating
functional for proper vertices) has the same symmetries as the action if
the regularization procedure respects the symmetries. This is not the
case for non-linear realizations for which the radiative corrections
change the form of the symmetry transformation. On top of this, in the
linear realizations, the symmetries manifest themselves at the level
of the effective action as linear Ward identities, while one needs to
invoke more involved Slavnov-Taylor identities in the non-linear case.

Nonlinear realizations of symmetry groups appear in very different
physical situations, both in condensed matter and particle
physics. They play a major role, for example, for the analysis of the
low momentum behavior of theories where some symmetry is spontaneously
broken (as in magnetic systems in the low-temperature phase or in
strong interactions physics where chiral symmetry is spontaneously
broken). They appear also in the formulation of many supersymmetric
models (see for instance \cite{Weinberg00}). A third example of
symmetry realized non-linearly take place in Quantum Chromodynamics
(QCD). For this theory, most of the theoretical approaches use the
Fadeev-Popov procedure to fix the gauge. In a seminal work, Becchi,
Rouet, Stora \cite{Becchi74,Becchi75} (and independently Tyutin
\cite{Tyutin75}) observed that the resulting theory presents a
symmetry, referred to as BRST symmetry. Today, BRST symmetry is a
cornerstone of our understanding of nonabelian gauge theories. It
simplifies considerably the proofs of renormalizability and unitarity
of the theory. Moreover, the Fadeev-Popov procedure enlarges the ket
space and the BRST symmetry is fundamental to reduce it to the
physical Hilbert space. As we shall discuss in more details below, the
BRST transformation is nonlinear: some fields have variations that are
quadratic.

There is, however, a point where BRST is at odds with most nonlinearly
realized symmetries. In all other cases one knows how to construct a
model where the symmetry is realized linearly and that behaves in the
same way (at least at large distances) as the non-linearly realized
one. This is the case for any continuous bosonic internal symmetry
\cite{Palais57,Mostow57}, and is also the case for standard
supersymmetries \cite{Weinberg00}. However, the construction of such a
model was missing for the BRST symmetry and this is the subject of the
present letter.

In all cases where such a linear realization of the symmetries is
available, it proves to be extremely clarifying on the conceptual
level and very useful in practical calculations. For example, when the
nonlinear symmetry takes its origin in a spontaneously broken
symmetry, the linear realization is essential in order to understand the
{\it restoration} of the symmetry induced by fluctuations. Moreover,
even if nonlinear realization of symmetries can be handled at a
perturbative level, they become intractable nonperturbatively. 

These difficulties are particularly problematic for two functional
approaches that try to formulate approximation schemes that go beyond
perturbation theory: the Non-Perturbative Renormalization Group (NPRG)
\cite{Wetterich92,Berges00} and the 2-PI formalism
\cite{Cornwall74,Berges00b,Berges04}. Both approaches rely on
introducing in the action a term quadratic in the fields that is
interpreted as a regularizing term in the framework of the NPRG and as
a source for composite operators in the 2-PI formalism. The variation
of this quadratic term under non-linear symmetries is not quadratic
itself so this term explicitly breaks the symmetry. Actually, in both
approaches, the regularizing function/sources are put to zero at the
end of the calculation so one could, at least in principle
\cite{Becchi03,Ellwanger94}, recover the symmetry in the 'physical'
limit, but this requires in general to impose fine-tuning conditions.

In many cases, this difficulty has been avoided with great success by
exploiting the equivalence of such a model with another where the
symmetry is realized linearly \cite{Berges00}. However, in absence of
a linear realization of the BRST symmetry for Yang-Mills theory, this
strategy has been beyond reach for QCD. The
construction of a linear realization of BRST which is addressed in
this letter is then an interesting starting point for formulating
functional methods for QCD where the BRST symmetry is respected at all
intermediate steps.

The letter is organized as follows. In Sect. 2 we discuss the linear
and nonlinear versions of a simple bosonic model. This is used to
illustrate the ideas developed in the following of the article. In
Sect. 3, we briefly present the Curci-Ferrari-Delbourgo-Jarvis (CFDJ)
gauge-fixing of Yang-Mills (YM) theory. The associated gauge-fixed
action has a large group of symmetries which simplify our
discussion. In Sect. 4 we propose a model which is shown to coincide
with the YM theory in the CFDJ gauge fixing at large distances but for
which the BRST symmetry is realized linearly. We give our conclusions
in Sect. 5.

\section{The case of the \nls{}}

Before discussing the construction of a linear realization of BRST
symmetry, we will review first a similar construction for a much
simpler field theory: the Nonlinear $\sigma$ (\nls{}) model in two
dimensions. As is well known, Quantum Chromodynamics (QCD) shares many
illuminating similarities with it that go beyond the nonlinear
realization of symmetries. To cite the most striking similarities,
both theories present asymptotic freedom: the effective coupling
constant vanishes logarithmically at high momentum scales. In the
infrared regime, both theories present a scale at which the spectrum
changes drastically, this scale being generically (without fine tuning
of the microscopic parameters) much smaller than the relevant
ultraviolet scales. For QCD, this momentum scale is
$\Lambda_{\mathrm{QCD}}\sim 200$~MeV which is 19 orders of magnitude
smaller than the Planck scale. For momenta larger than
$\Lambda_{\mathrm{QCD}}$, the relevant degrees of freedom are quarks
and massless gluons, while the low energy spectrum has a gap and is
made of hadrons. The \nls{} model presents a similar property. The
infrared scale is now $\Lambda_{\mathrm{NL}\sigma}$ which is
generically found to be orders of magnitude smaller than the inverse
lattice spacing. For higher momentum scales, the theory is described
in terms of massless Goldstone modes while, in the opposite regime,
the spectrum presents massive modes. A surprising property of the
\nls{} model is that the number of Goldstone modes in the ultraviolet
does not coincide with the number of massive modes in the infrared.

Let us describe in more details this model in the case of the $O(N)\to
O(N-1)$ symmetry-breaking scheme. It is parametrized in terms of a
$N-1$ component field $\pi_i$ ($i=1,\dots, N-1$) that interacts via a
euclidean action
\begin{equation}
\label{eq:nls}
S=\frac 1{2g^2}\int d^2x\ \left((\nabla \pi_i)^2+\frac{(\pi_i \nabla
  \pi_i)^2}{1-\pi^2}\right).
\end{equation}
On top of the O($N-1$) linear symmetry group that consists in rotating
the $\pi$'s ($\pi_i\to\pi_i+\epsilon_{ij} \pi_j$ with $\epsilon_{ij}$
antisymmetric), this action is invariant under the non-linear
transformations: $\pi_i\to\pi_i+\epsilon_i \sqrt{1-\pi^2}$. These
$N-1$ symmetries together with the linearly realized ones generate the
O$(N)$-symmetry group of the action $S$.

Nonlinear symmetries translate into Slavnov-Taylor
\cite{Slavnov72,Taylor71} equations that are hard to handle because
they are nonlinear in the effective action $\Gamma$.  A crucial
observation is that it is possible to construct a theory which is
equivalent to the \nls{} model at large distances, but in which all
symmetries are realized linearly. To do so, one adds-up a massive
field $\sigma$ that completes the standard $N$-multiplet of the linear
$O(N)$ model. The action is the standard Landau-Ginsburg one
\begin{equation}
  \label{eq:LG}
  S= \frac1{2g^2}\int d^2x\  [ (\nabla \sigma)^2+(\nabla
  \pi_i)^2+M^2(\sigma^2+\pi_i^2-1)^2].
\end{equation}
In the limit of low momenta (that is for momenta much smaller than
$M$), the $\sigma$ mode is frozen. The potential term can be seen as a
constraint for the field $\sigma$, which can therefore be replaced by
the solution of its equation of motion, that is
$\sigma\to\sqrt{1-\pi_i^2}$. One then recovers the non-linear version
of the model~(\ref{eq:nls}).

The linear realization of the symmetry
has many virtues. At a technical level, the linearly-realized
symmetries impose Ward identities (that are linear in $\Gamma$) which
are much simpler to handle than Slavnov-Taylor identities in actual
calculations.  Another virtue, which is probably more important, is
that it helps understanding the infrared sector of the theory and the
way in which the spectrum of the theory changes at the scale
$\Lambda_{\mathrm{NL}\sigma}$. Upon renormalization, the minimum of the
effective potential is shifted toward the origin. At a scale of the
order of $\Lambda_{\mathrm{NL}\sigma}$ and for all smaller momentum
scales, the effective potential has actually a single minimum at the
origin. The symmetry which was spontaneously broken in the ultraviolet
is restored by fluctuations. Consequently, instead of having $N-1$
massless Goldstone bosons as one could naively guess from
Eq.(\ref{eq:nls}), the spectrum is actually composed of $N$ degenerate
massive fields. It is interesting to note that
$\Lambda_{\mathrm{NL}\sigma}\sim M \exp(-\mathrm{cte}/g^2)$ which shows that one need
not perform a fine-tuning of the parameters to have a UV
scale $M$ much larger than $\Lambda_{\mathrm{NL}\sigma}$.

The strong analogies between QCD and the \nls{} model call for a
parallel treatment of these theories \footnote{We should stress,
  however, a strong difference between the two models: there is no
  equivalent in the \nls{} model of the confinement phenomenon that
  shows up in QCD.}.  In particular, since the linear realization of
the symmetry in the \nls{} model is so useful for understanding its
infrared regime, one can expect that a similar
construction for QCD would give a clarifying viewpoint on this theory
in the infrared sector. 

\section{Yang-Mills theory in Curci-Ferrari-Delbourgo-Jarvis gauge}

In the next sections, we will present a model where such a linear
realization takes place. We will show that it behaves at large distances
as the YM theory in the CFDJ gauge fixing
\cite{Curci76,Delbourgo81}. This class of gauge fixing admits the
Landau gauge as a particular case. Moreover, as we recall later, it
presents a large symmetry group that considerably simplifies the
analysis. We show that, very much as for the \nls{} model, one can add
massive fields (of typical mass $M$) to those of the gauge-fixed YM
theory (called light fields in the following) such that the BRST
symmetry is realized linearly.  In the low-energy limit, where the
massive fields decouple, one recovers the YM theory for the light
fields.

Before presenting that model, it is convenient to describe first the CFDJ gauge fixing and its symmetries. The action in the euclidean space reads:
\begin{equation}
\label{eq:lagdiv}
S=\int d^dx\ (\mathcal{L}_{\rm{YM}}+\mathcal{L}_{\rm{GF}}).
\end{equation}
$\mathcal{L}_{\rm{YM}}$ is the YM Lagrangian:
\begin{equation}
\label{eq:lagYM}
\mathcal{L}_{\rm{YM}}=\frac{1}{4}F_{\mu\nu}^\alpha F_{\mu\nu}^\alpha,
\end{equation}
$F_{\mu\nu}^\alpha=\partial_\mu A_\nu^\alpha-\partial_\nu A_\mu^\alpha+g
f^{\alpha\beta\gamma}A_\mu^\beta A_\nu^\gamma$ is the field strength, $g$ is the
gauge coupling, $A_\mu$ is the gauge field, and $f^{\alpha\beta\gamma}$ denotes the
structure constants of the gauge group that are chosen completely
antisymmetric.  $\mathcal{L}_{\rm{GF}}$ is the gauge fixing term, which
includes a ghost sector. It takes the form:
\begin{equation}
\label{eq:lagcf}
\mathcal{L}_{\rm{GF}}=
\frac{1}{2}\partial_\mu \bar c^\alpha (D_\mu c)^\alpha
+\frac{1}{2}(D_\mu \bar c)^\alpha \partial_\mu c^\alpha+\frac{\xi}{2}h^\alpha h^\alpha
+ih^\alpha\partial_\mu A_\mu^\alpha 
-\xi\frac{g^2}{8}(f^{\alpha \beta \gamma}\bar c^\beta c^\gamma)^2.
\end{equation}
Here, $c$ and $\bar c$ are ghost and antighosts fields respectively,
$h$ is the Lagrange multiplier field
and $(D_\mu \varphi)^\alpha= \partial_\mu \varphi^\alpha + g f^{\alpha \beta\gamma} A_\mu^\beta
\varphi^\gamma$ is the covariant derivative for any field $\varphi$ in the
adjoint representation. 

The gauge fixing Lagrangian is invariant under a) the euclidean
symmetries of the spacetime; b) the global color symmetry; c) the
ghost conjugation symmetry: $c^\alpha \to \bar c^\alpha$,
$\bar c^\alpha \to -c^\alpha$;
d) the continuous symplectic group $SP(2,\mathbb{R})$
\cite{Delduc89} with generators $N$, $t$ and $\bar t$ that act only on
the ghost sector, and defined by:
\begin{eqnarray}
\label{symmsp2}
t c^\alpha&=\bar t\bar c^\alpha= 0  \nonumber \\
t \bar c^\alpha&= N c^\alpha  = c^\alpha \nonumber \\
\bar t c^\alpha&=  N \bar c^\alpha= - \bar c^\alpha;
\end{eqnarray}
e) the model is also invariant under the {\em nonlinear}
BRST symmetry:
\begin{eqnarray}
\label{symmbrst}
&s A_\mu^\alpha=(D_\mu c)^\alpha\nonumber,\\
&s c^\alpha= -\frac{g}{2} f^{\alpha\beta\gamma} c^\beta c^\gamma,\nonumber\\
&s \bar c^\alpha =ih^\alpha -\frac{g}{2} f^{\alpha\beta\gamma}\bar c^\beta c^\gamma,\nonumber\\
&s\, i h^\alpha=\frac{g}{2} f^{\alpha\beta\gamma} \Big(i h^\beta c^\gamma+\frac{g}{4} f^{\gamma \delta\epsilon} \bar
c^\beta c^\delta c^\epsilon\Big).
\end{eqnarray}
By virtue of the ghost conjugation symmetry, the action is also
invariant under an anti-BRST symmetry $\bar s$.  These symmetries
satisfy the standard nilpotency property ($s^2=\bar s^2=\bar s s+s\bar
s=0$); f) recently, we showed that the CFDJ action is also invariant
under four gauge supersymmetries \cite{Tissier08}. One corresponds to
a local shift of the field $h$. A second one corresponds to a shift of
the ghost field by an infinitesimal local parameter simultaneously
with a color transformation of the field $h$ with the same
parameter. A third one is just the ghost conjugated of the previous
one. The last one corresponds to a gauge transformation of the
fields. These gauge transformations are not truly symmetries but the
associated variations of the action are linear in the fields so that
one can write a simple Ward identity associated with them, that in
particular imply some non renormalization theorems
\cite{Tissier08,Wschebor07}.

\section{Realizing linearly the BRST symmetry}
Having described the symmetries of the CFDJ gauge fixing, let us now
present a model in which those symmetries are realized linearly. The arena of
the model is a superspace with $d$ bosonic coordinates $x^\mu$ and
two grassmanian anticommuting coordinates $\ts$ and $\tb$
($\ts^2=\tb^2=\ts\tb+\tb\ts=0$).  It is convenient to work in such a
superspace because, as well known since the work of Tonin and Bonora
\cite{Bonora80,Bonora81} (see also \cite{Baulieu81,Delbourgo81}),
the BRST symmetry appears then as an invariance under translation in
the $\tb$ direction. Similarly the translation in the $\ts$ direction
yields the symmetry $\bar s$. The geometry of the superspace is now
dictated by the symmetries of the CFDJ action. A first guess would be to
take a flat superspace. However this superspace admits
superrotations that mix the bosonic and fermionic coordinates, which have no
equivalent in the CFDJ action. In
what follows, we therefore choose the geometry of the superspace in
such a way that these unwanted symmetries are explicitly broken. To
implement this idea, we consider a Riemannian superspace with metric:
\begin{equation}
g_{AB}=
\left\{
\begin{array}{ll}
\delta_{\mu\nu}\hspace{.5cm} &\rm{if\ } A=\mu,\ B=\nu\\
-(1+M^2\tb\ts)& \rm{if\ } A=\ts,\ B=\tb\\
(1+M^2\tb\ts)& \rm{if\ } A=\tb,\ B=\ts\\
0& \rm{otherwise.}
\end{array}
\right.
\end{equation}
Here and below, the uppercase indices run over bosonic and fermionic
components of the superspace. The formulas of the Riemannian geometry
in the superspace are essentially identical to those in bosonic
spaces, except for signs as explained in \cite{Nath75,Arnowitt75}.
Here, contrarily to \cite{Nath75,Arnowitt75} we use left derivatives.

The isometries of this superspace have been discussed in details in
\cite{Tissier08}. Let us here comment on some of its properties. This
curved superspace has a typical length scale $M^{-1}$.  At larger
length scales, the grassmannian directions are wrapped around and the
space is similar to a purely bosonic space. In the other limit, at
length scales much smaller than $M^{-1}$, the curvature is irrelevant
and the space is equivalent to a flat superspace with $d$ bosonic and
2 fermionic directions. As is well known since the pioneering work of
Parisi and Sourlas \cite{Parisi79} (see also \cite{McClain82}),
theories defined in this kind of flat superspace present the property
of dimensional reduction. This means that a theory in this superspace
has the same correlation functions as the equivalent theory in a
purely bosonic space with $d-2$ dimensions if their isometries are not
spontaneously broken. For this reason, it is
often said that the grassmanian directions count as negative
dimensions. In a sense, the superspace considered here realizes the
standard idea of adding extra confined dimensions of space as possible
sources of new physics. However in the case considered here, the
wrapped directions are fermionic so that there are a finite number of
excitations associated with these extra dimensions \cite{Delbourgo07}.

Having described the superspace and its isometries, let us now come to
the field content of the theory. We consider a vectorial superfield
$\mathcal A^A=\{\mathcal A^\mu,\mathcal A^\theta,\mathcal A^\tb\}$ in this space where the
last two components are fermionic. The action for this field is chosen to
respect the isometries of the superspace. We therefore
contract the superspace indices with the tensors associated with the
space, that are in our case the metric, the Riemann and the Ricci
tensors. In
the superspace we are considering, the Riemann tensor can be expressed
in terms of the Ricci tensor so we do not consider it as an
independent tensor in what follows. Moreover, the Ricci tensor
$R_{AB}$ is equal to $M^2g_{AB}$ in the fermionic sector, and zero in
the bosonic one. In order to satisfy the isometries of the superspace,
it is, as usual, necessary to use an invariant measure $\int_x=\int
\sqrt{\mathrm{sdet}\ g}\ d^dxd\ts d\tb$ \cite{Arnowitt75} (where sdet is the superdeterminant). There is
obviously an ambiguity in the sign of the square root. For later
convenience, we choose $\sqrt{\mathrm{sdet}\ g}=-1+M^2\tb\ts$.

As discussed above and more thoroughly in
\cite{Tissier08}, the CFDJ presents a gauge symmetry up to some non renormalized terms. 
We therefore require the theory to be
invariant under the gauge transformation $\mathcal A_A^\alpha\to\mathcal
A_A^\alpha+\partial_A\Lambda^\alpha +gf^{\alpha \beta \gamma}\mathcal A_A^\beta \Lambda^\gamma$ with
$\Lambda^\alpha=\Lambda^\alpha(x,\theta,\bar\theta)$ an arbitrary function of the superspace coordinates. In the
following, in order to respect the gauge invariance, we write the
action in terms of the field strength
\begin{equation}
\mathcal F_{AB}^\alpha=(-1)^b(\partial_A\mathcal
A_B^\alpha-(-1)^{ab}\partial_B\mathcal A_A^\alpha+gf^{\alpha \beta \gamma} \mathcal A_A^\beta
\mathcal A_B^\gamma)
\end{equation}
where the lowercase letters are 1 if the associated uppercase letters
are fermionic, and 0 otherwise. The action therefore
has a Yang-Mills-like term
\begin{equation}
\mathcal S_{\mathrm{YM}}=-\frac{1}4 \int_x(-1)^a\mathcal
F_{AB}g^{BC}\mathcal F_{CD}g^{DA}
\end{equation}
which is clearly gauge invariant and invariant under the isometry
group, thanks to the $(-1)^a$. Of course other terms with the same
symmetries can be constructed but we only consider the simplest one
here.

If we had only this term in the action, we would be in deep trouble
since we could not invert the 2-point function to build the
propagator. Fortunately it is possible, as in QED, to add a mass term
in the theory that breaks gauge invariance in a controlled way.
Actually, because of the structure of the curved superspace there are
two independent mass-like terms invariant under the isometries of the
space. We therefore add two mass terms to the action:
\begin{equation}
\label{massterm}
\mathcal S_{\mathrm{mass}}=\int_x\left(\frac{m^2}2g_{AB}
+\frac{\nu}2 R_{AB}\right)\mathcal A^A\mathcal A^B.
\end{equation}

The mass term breaks the super-gauge invariance in a simple
way. Actually its variation under gauge transformation is linear in
the fields so one can write a linear Ward identity for this
transformation, that reads:
\begin{equation}
D_B\left(g^{AB} \frac{\delta (\Gamma-S)}{\delta \mathcal A^A_\alpha}\right
)=0.
\end{equation}
This identity directly shows that the part of the action that breaks
gauge-invariance is not renormalized.  The breaking of the gauge
invariance induced by these mass terms has a counterpart in the CFDJ
model. As mentioned in the previous section (see discussion of the
symmetry f) above) the CFDJ action is not invariant under the
super-gauge transformation but the variation of the action under this
transformation is also linear in the fields.

Let us now discuss how the decoupling of the massive modes works. To
do so, we expand the field components in a Taylor series in the
grassmannian coordinates $\ts$ and $\tb$:
\begin{eqnarray}
  \mathcal A_\mu^\alpha(x,\theta,\bar\theta)&=A^\alpha_ \mu(x) + \tb B^\alpha_\mu(x) - \ts {\bar B}^\alpha_\mu (x)+ \tb
  \ts E^\alpha_\mu(x), \nonumber \\
\mathcal A_\ts^\alpha(x,\theta,\bar\theta) &=- \bar c^\alpha(x) + \ts \bar d^\alpha(x) - \tb
  b^\alpha(x) - \tb \ts \bar F^\alpha(x), \nonumber \\
\mathcal A_\tb^\alpha(x,\theta,\bar\theta) &= c^\alpha(x) + \tb d^\alpha(x) + \ts
  \bar b^\alpha(x) + \tb \ts F^\alpha(x).
\end{eqnarray}
The fields $A$, $B$, $\bar B$, etc. are standard fields in the $d$-dimensional euclidean
space. As we now show, most of them are massive, and therefore are decoupled in the
infrared.
It is straightforward to write down the action $\mathcal
S_{\mathrm{YM}}+\mathcal S_{\mathrm{mass}}$ in terms of these fields. This
leads to a lengthy expression that need not be given here. We just
reproduce the leading order of $\mathcal S_{\mathrm{YM}}$ at large
$M^2$, which reads
\begin{eqnarray}
\mathcal S_{\mathrm{YM}}^{\mathrm{large} M}&=M^2\int d^dx\Big[\frac 32(b^\alpha-\bar b^\alpha+
  gf^{\alpha\beta\gamma}\cb^\beta c^\gamma)^2-6(\bar d^\alpha+\frac
  g2f^{\alpha\beta\gamma}\cb^\beta\cb^\gamma)(d^\alpha+\frac
  g2f^{\alpha\beta\gamma}c^\beta c^\gamma) \nonumber \\
&+2(\bar B_\mu^\alpha-D_\mu \cb^\alpha)(B_\mu^\alpha-D_\mu
  c^\alpha)+\frac 14
  (F_{\mu\nu}^\alpha)^2\Big]
\end{eqnarray}
The relevant regime of parameters and momenta that allows us to
recover the usual YM theory is $M\gg p\gg m$. The first three terms
play the role of constraints for the fields $b-\bar b$, $B_\mu$, $\bar
B_\mu$, $d$, $\bar d$, which can be replaced in the rest of the
expression by their classical values $-
gf^{\alpha\beta\gamma}\cb^\beta c^\gamma$, $D_\mu c$, $D_\mu \cb$, $-\frac
g2f^{\alpha\beta\gamma}c^\beta c^\gamma$ and $-\frac
g2f^{\alpha\beta\gamma}\cb^\beta \cb^\gamma$ respectively. One then
obtains an action for the remaining fields that is actually quadratic
in $E_\mu$, $F$ and $\bar F$. One can then integrate over these fields
and get an action for $A_\mu$, $c$, $\bar c$ and $b+\bar b$. These
are, up to renormalization factors, the 4 light fields that appear in
the gauge-fixed CFDJ action. It is important to note that the
equations of motion for the massive modes give the same expression as
the {\em transversality conditions} in
\cite{Bonora80,Bonora81,Baulieu81,Delbourgo81}. Eliminating the
massive modes is therefore equivalent to imposing these transversality
conditions, except for the important fact that these relations need
not be imposed externally here.  Actually, in order to retrieve the
Lagrangian (\ref{eq:lagdiv},\ref{eq:lagYM},\ref{eq:lagcf}), one has to
make the replacements $A_\mu\to A_\mu/ M$, $c\to c/ m$, $g\to g M$,
$b+\bar b \to -2i h M/m^2$ and $\xi=(m^2+M^2\nu)M^2/m^4$. This
concludes the proof that the gauge theory in the curved superspace is
equivalent to the CFDJ gauged-fixed theory in the low energy limit. In
fact, it is interesting to note that in the context of the
transversality conditions formalism it has already been noted
\cite{Delbourgo81} that the gauge-fixing term is obtained by imposing
the transversality conditions to the mass term (\ref{massterm}).

\section{Conclusions and perspectives}

To conclude, we have presented a model in which the BRST symmetries
are realized linearly and which reduces to the YM theory in the CFDJ
gauge at long distances. This model treats on an equal footing the
gauge and ghosts fields. At distances much smaller than the inverse of
the mass of some massive fields, the model behaves as the massive YM
theory in two dimensions, which is renormalizable (see for example
\cite{Zinn-Justin}).

With a model where BRST symmetry is realized
linearly, we are for the YM theory in a position very similar to that
of the $\sigma$ model. We recall that in this case, the existence of the linear
version of the $\sigma$ model plays a fundamental role in order to exploit
functional methods (non-perturbative renormalization group equations, and 2-PI
formalism) respecting in each steps all the symmetries of the model. It is easy
to show that the present model gives a 2-PI functional respecting all the
symmetries of the model in a linear way. In what concerns the non-perturbative
renormalization group formalism, things are more involved. All isometries
(including the BRST symmetry) are respected trivially along the flow but the
control of the supergauge symmetries requires further analysis. This
work is in progress.

Another aspect where the linear version of the model is helpful for the $\sigma$
model is in clarifying the appearance of the mass
gap and we can wonder if the present model has the same virtue for YM
theory. Of course, we have no definitive answer, but we speculate that
the infrared fluctuations effectively flatten the superspace. If this
is the case, at distances much larger than
$\Lambda_{\mathrm{QCD}}^{-1}$, all fields should be treated on an equal
footing. Thanks to the dimensional reduction property
\cite{Parisi79,McClain82}, the theory would then behave in the
infrared as YM in two dimensions, which is expected to have a physical
mass gap, at least for large $N$ \cite{tHooft73}.

This work opens the way to several developments. In light of the
previous discussion, it is important to study the loop corrections to
the present analysis. One should also determine how to reduce the
state space to the physical Hilbert space, probably in a similar way
as in the standard BRST formalism. It would also be interesting to
investigate if similar constructions can be made to recover at low
energy other gauge fixings and to introduce matter fields. Moreover,
in this Letter, we focused on the case of YM theory, but we plan to
generalize the present work to other gauge theories, in particular to
gravity. If what was done here translates to that case, one would
recover in the ultraviolet the quantum gravity in 2 dimensions
\cite{Ambjorn05,Lauscher05}, which is known to be renormalizable.

\section{Acknowledgments}
M. T. thanks the IFFI for hospitality, where this work was done. We
thank J.-P. Blaizot, B. Delamotte, M. Reisenberger and G. Tarjus for useful
discussions. We acknowledge support of PEDECIBA, the ECOS program and
PDT uruguayan program.  LPTMC is UMR 7600 of CNRS.

\bibliographystyle{unsrt}

\end{document}